\documentclass[conference]{IEEEtran}
\usepackage{cite}

 \usepackage{amsbsy}

\ifCLASSINFOpdf
\else
\fi

\usepackage{graphicx}

\usepackage{url}

\begin{document}
%
\title{Identifying Similar Patients Using Self-Organising Maps: A Case Study on Type-1 Diabetes Self-care Survey Responses}

\author{\IEEEauthorblockN{Santosh Tirunagari\IEEEauthorrefmark{1},
Norman Poh\IEEEauthorrefmark{2},  Guosheng Hu\IEEEauthorrefmark{3} and
David Windridge\IEEEauthorrefmark{4}}
\\\IEEEauthorblockA{University of Surrey, Guildford, Surrey GU2 7XH\\
Email: \IEEEauthorrefmark{1}s.tirunagari@surrey.ac.uk,
\IEEEauthorrefmark{2}n.poh@surrey.ac.uk,
\IEEEauthorrefmark{3}g.hu@surrey.ac.uk,
\IEEEauthorrefmark{4}d.windridge@surrey.ac.uk}}


\maketitle

\begin{abstract}
Diabetes is considered  a lifestyle disease and a well managed self-care plays an important role in the treatment. Clinicians often conduct surveys to understand the self-care behaviours in their patients. 
In this context, we propose to use Self-Organising Maps (SOM) to explore the survey data  for assessing the self-care behaviours in Type-1 diabetic patients. Specifically, SOM is used to visualise high dimensional similar patient profiles, which is rarely discussed. Experiments demonstrate that our findings through SOM analysis corresponds well to the expectations of the clinicians. In addition, our findings inspire the experts to improve their understanding of the self-care behaviours for their patients. The principle findings in our study show: 1) patients who take correct dose of insulin, inject insulin at the right time, 2) patients who take correct food portions undertake regular physical activity and 3) patients who eat on time take correct food portions.

\end{abstract}


%
\IEEEpeerreviewmaketitle

\section{Introduction}

Diabetes is a major health problem in the present world. It is caused due to the presence of excess glucose within the blood. It is of two kinds: 1) Type-1 and 2) Type-2. Type-1 diabetes is a condition where the immune cells attack the insulin manufacturing cells. As a result, patients with Type-1 diabetes cannot produce insulin and need insulin injections to survive. It is estimated that Type-1 diabetes constitutes 10\% to 15\% of the total diabetic patients in the world. The Type-1 diabetes appears usually before 40 years of age. It is the most common type of diabetes found in the children~\cite{frayling2007common}. Type-1 diabetes is typically represented as a `lifestyle disease', as a result it is common in patients: 1) who do not perform enough physical activity 2) who are obese (having excess body weight) and 3) have high cholesterol. Hence, to effectively manage this disease we consider the following self-care factors. 

\begin{itemize}
\item Check blood glucose with monitor.
\item  Record blood glucose results.
\item  Check ketones when glucose level is high.
\item  Take correct dose of insulin. 
\item  Take insulin at the right time.
\item  Eat the correct food portions. 
\item  Eat meals/snacks on time. 
\item Keep food records.
\item Read food labels. 
\item  Treat low blood glucose with just the recommended amount of carbohydrate. 
\item  Carry quick acting sugar to treat low blood glucose. 
\item Come in for clinic appointments. 
\item  Wear a medic alert ID. 
\item  Exercise. 
\item  Adjust insulin dosage based on glucose values, food, and exercise.
\end{itemize}

Daily insulin injections, healthy diet and regular physical activity are the most important factors that influence the treatment for this disease. Therefore, we in this study aim to analyse these self-care factors using statistical machine learning methods.

\subsection{Motivation for using SOM}

Health care survey data contain a wealth of information about the patient behaviours. Identifying similar patients with poor self-care behaviours helps clinicians in suggesting suitable treatment plans.
The survey data are usually heterogeneous and contain missing values. Applying traditional statistical methods might give clinicians overly simplified conclusions. Therefore, there is a need for statistical machine-learning tools such as SOM for the data analysis. SOM algorithm is a well applied data analysis technique for mining correlations and clustering similar responses in the surveys. The clustered responses in the higher dimensions are mapped onto a 2-dimensional grid for visualisation. This makes SOM a suitable method in visualising high-dimensional data by revealing more meaningful insights.

Previously, SOM has been used to visually explore data in the datasets such as: financial~\cite{deboeck1998visual}, gene expression~\cite{tamayo1999interpreting} and linguistics~\cite{honkela1997websom}. A paper by Toronen $et$ $al.$~\cite{toronen1999analysis} has used SOM algorithm to analyse yeast gene expression data and showed that SOM is an excellent tool for the analysing and visualising gene expression profiles. Mehmood $et$ $al$.~\cite{mehmood2011self} found interesting correlations between countries and nutrition, health and lifestyle using SOM analysis. Tirunagari $et$ $al.$ used SOM analysis to correlate causal factors associated with different types of maritime accidents in ~\cite{tirunagari2012mining} as well as identifying patient profiles who are in need of selfcare management in~\cite{tirunagari2014patient}. Lagus $et$ $al.$~\cite{lagus2013exploration} applied SOM in analysing loneliness questionnaire data to identify profiles of loneliness for improving the social well being among the different sub groups.

\subsection{Objectives}
The main objectives of this study are two-fold. Firstly, from the computational perspective, we would like to examine the feasibility of using SOM as a means of extracting  useful information from survey questionnaires. Secondly, from the scientific perspective, we would like to understand if the responses collected from the Type-1 Diabetes survey are reasonable and correspond to what domain experts and clinicians would expect. For instance, is it possible to  answer the following questions:
\begin{itemize}
\item How are the individual patients grouped together? 
\item How are the self-care factors correlated?
\end{itemize}

\subsection{Contributions}
Our contributions can thus be summarised as follows:
\begin{itemize}
\item {\bf Novel use of SOM for visualising similar patients from the data}~ Although SOM has been widely used in exploring the survey data, its uses in visualising  similar patient profiles are rarely discussed. Our approach of visualising similar patient profiles will prove to be useful, as concurred by the domain experts.
\item {\bf Improved understanding}~ We will demonstrate that the visual analytics provided by SOM can improve experts' understanding of the impact of self-care behaviours of patients with Type-1 diabetes. 
\end{itemize}
Therefore, SOM is a potentially viable tool for analysing high-dimensional questionnaire responses as well as a means for visually summarising individual patients data.

\subsection{Organisation}
 The organisation of the paper is as follows: In section~\ref{data}, we present the survey dataset. In section~\ref{methods}, we present method SOM. Experiments and results are discussed in section~\ref{expres} in cluing data preprocessing and imputation (filling out the missing values). Finally, in section~\ref{condis}, we draw conclusions and summarise with the discussions.

\section{Dataset}
\label{data}

\begin{figure}[ht]
\centering
\includegraphics[width=3.5in]{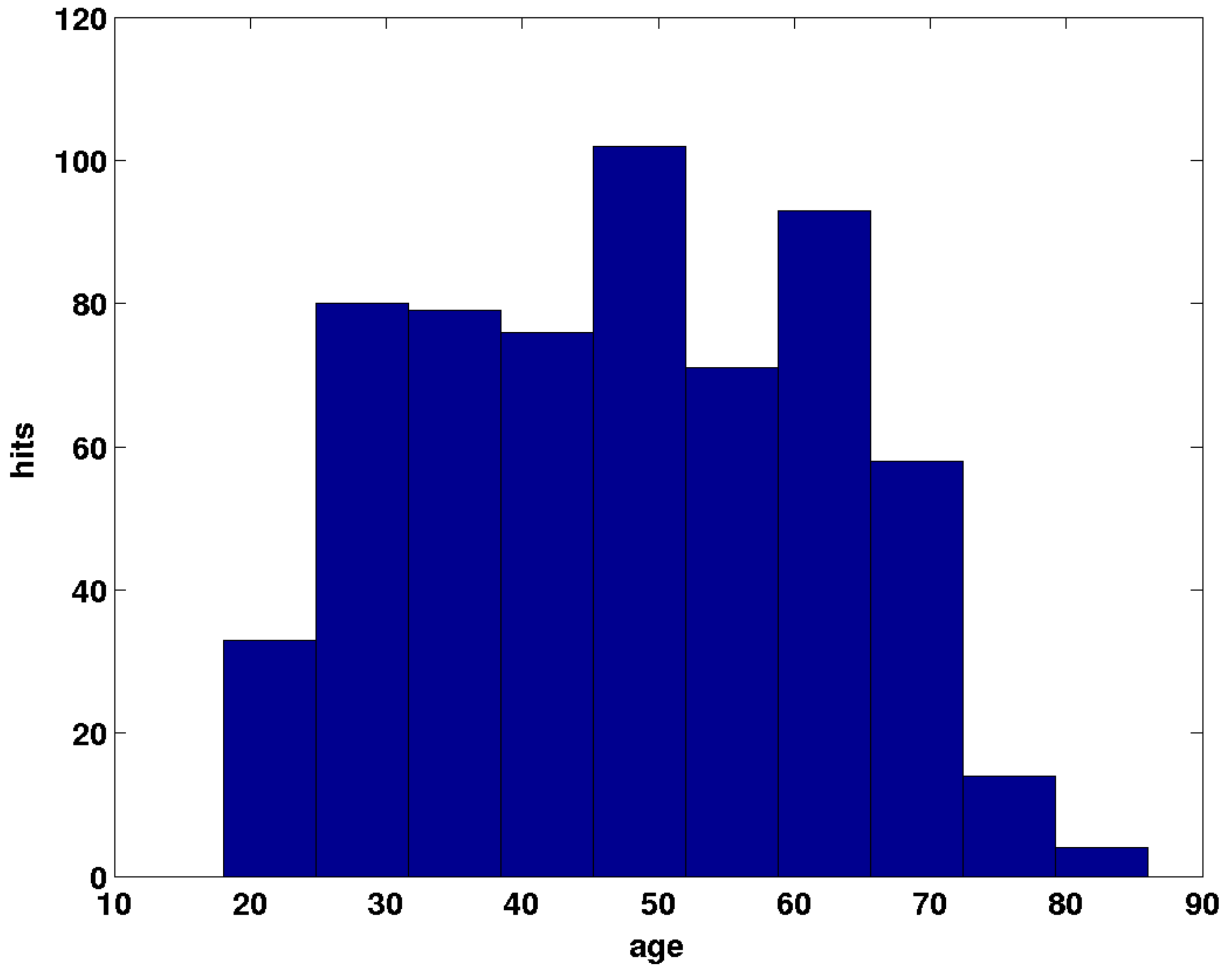}
\caption{Distribution of age with the majority of participants being between the ages of 25 and 70.}
\label{fig_age}
\end{figure}

The survey consists of 611 patients' responses for the 15 self-care behaviours~\cite{tirunagari2014patient}. The participants were all above 18 years of age with Type-1 diabetes. 
The responses for the self-care behaviours are one of the following:  1) Never 2) Rarely 3) Sometimes 4) Usually and 5) Always. The highest percentage of respondents is in their mid forties to early fifties as shown in Figure~\ref{fig_age}.

Male participants represent 45\% of the survey population while female respondents constitute 
55\%. The level of education has 91\% of respondents with some forms of qualification.

\section{Methods}
\label{methods}
To explore the survey dataset we have used self organising map (SOM). SOM also known as Kohonen map is an unsupervised technique that is most often described in the language of artificial neural networks. SOM provides a way of representing multidimensional data in two or three dimensions. This process of reducing the dimensionality of vectors  is based on a data compression technique known as vector quantisation. In addition, SOM creates a network that stores information in such a way that any topological relationships within the training set are maintained. Hence, SOMs are useful for visualising large data sets of high dimensionality. SOM is an unsupervised, competitive learning approach in which only one neuron wins each training phase. There are no connections between the neurons in input and output layer. However, they communicate with each other via a neighbourhood function. If a neuron wins in training phase, it will also impact its neighbours.

Let us consider the input vector
\begin{equation}
\boldsymbol{x}=[x_1,x_2, ..., x_n]^T
\end{equation}
The synaptic weight vector of the neuron $i$ in the output layer of 2-dimensional neurons is
\begin{equation}
\boldsymbol{w_i}=[w_i^1, w_i^2, ..., w_i^n]^T, i=1,2, \cdots, m;
\end{equation}
where, $m$ is the number of output neurons and $w_i^v$ is the weight associated with neuron $m$ and variable $v$. Although, the output
neurons are arranged in $2$-dimensional array, their weight vectors are $n$-dimensional i.e. same
dimensions as the input vector $\boldsymbol{x}$. The Euclidean distance $\|\boldsymbol{x}-\boldsymbol{w_i}\|_2$ of the current input vector
$\boldsymbol{x}$ to all the weight vector $i=1,2... m$ is computed. The winning neuron is one whose weight
vector $\boldsymbol{w_q}$ has the minimum Euclidean distance to $\boldsymbol{x}$, i.e.,
\begin{equation}
q(x)=arg_{min_i} \|\boldsymbol{x}-\boldsymbol{w_i}\|_2
\end{equation}
The weight vectors of the winning neurons and the neurons in its predefined neighbourhood
$\eta_q$ are updated using a gradient descent, leading to the following update rule:
\begin{equation}
 w_i (k + 1) = w_i (k) + \eta_{qi}(k)[x(k) - w_i (k)]
\end{equation}
Neurons outside the neighbourhood are not updated i.e $\eta_q(k)=0$ and neurons inside the neighbourhood  $\eta_q$ are updated using equation (\ref{eq1}):
\begin{equation}
\label{eq1}
\eta_{qi}(k)=\mu(k).
\end{equation} 
The learning parameter $\mu(k)$, where $0< \mu (k)<1 $ decreases with increasing iterations. The learning process has two phases: 1) ordering phase (rough training phase) and 2) convergence phase (fine training phase). In ordering phase, the topological ordering of the weight vectors is carried out. The learning parameter $\mu (k) $ is set close to unity. In convergence phase, the self-organising map is fine-tuned, which is achieved by setting the learning parameters $ \mu (k)$  in the order of 0.01. The stopping criterion for the SOM algorithm is the number of specified iterations or the smallest
degree of change in the weight vectors.

\section{Experiments and Results}
\label{expres}
In this section we identify similar patient profiles based on their self-care behaviours and present the results for factors relating to: 1) managing blood glucose 2) insulin intake 3) food and 4) physical activity. 

\subsection{Data preprocessing}

The data preprocessing included the conversion of string values into categorical numerical values for ease of computation. This conversion makes the data analysis computationally inexpensive. For `NA' results we used 0, `Always', `Often', `Sometimes', `Rarely' and `Never' are scored as 5, 4, 3, 2 and 1 respectively, and for NaNs the appropriate missing value using Matlab's `\emph{knnimpute(Data)}' function.

After the data preprocessing, 611 patient responses to 15 self-care factors are given as an input ($611-by-18$) to the SOM algorithm~\footnote{\url{http://www.cis.hut.fi/projects/somtoolbox/}}. The SOM program outputs a 2-dimensional visualisation map of patients groups across the considered self-care factors. In this experiment we have chosen 30$\times$30 grid for visualisation. The colorbar in the figures shows the user rating i.e 'Never' corresponding to darker blue, 'Rarely' to lighter blue, 'Sometimes' to green, 'Usually' to yellow and 'Always' to orange.

\subsection{Blood glucose}

\begin{figure}[h]
\centering
\includegraphics[width=3.5in]{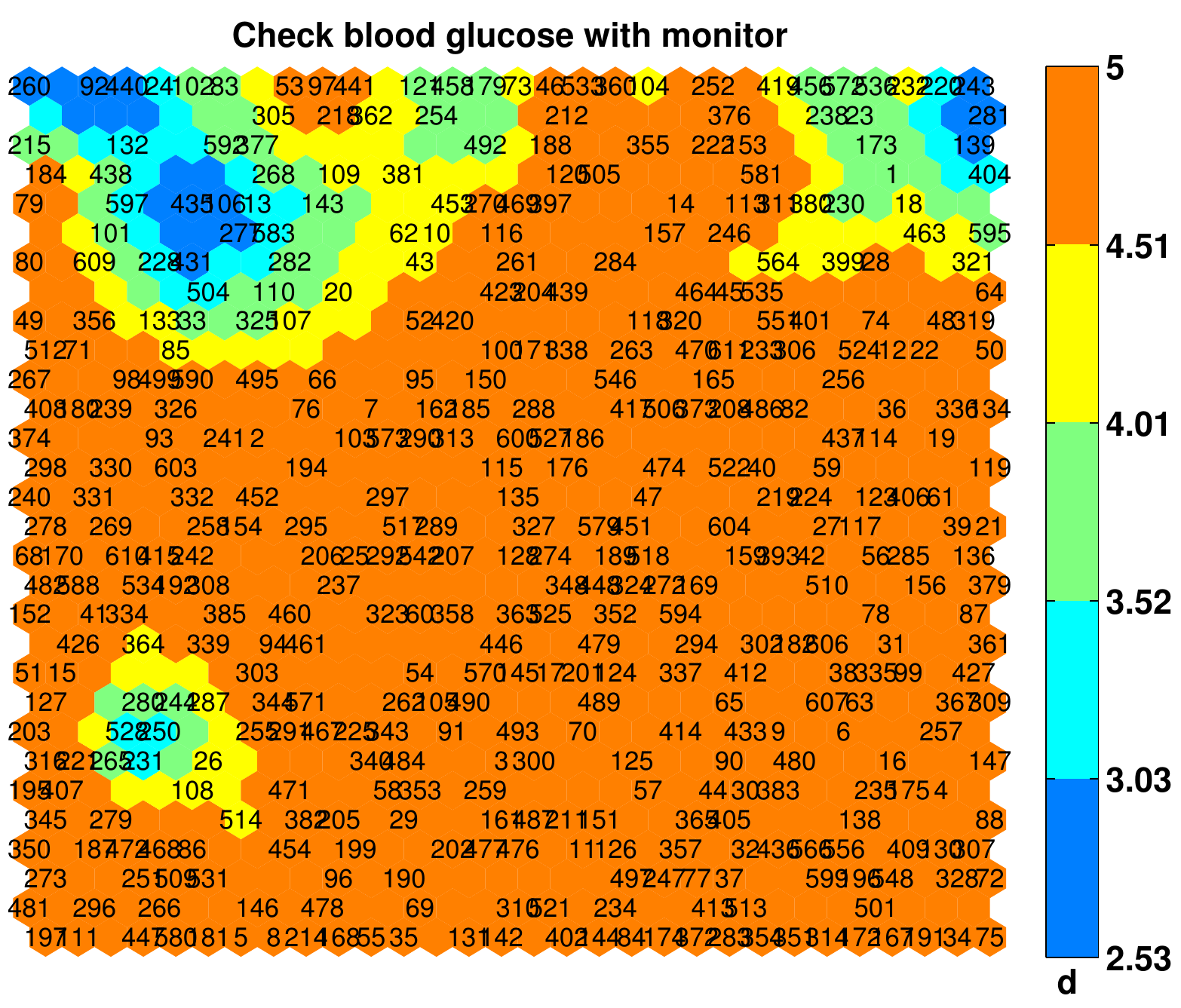}
\caption{Patient groupings according to the ratings given for the self-care factor: `Check blood glucose with monitor'.}
\label{fig_ps_1}
\end{figure}

Self-testing for blood sugar is very essential in managing a healthy diabetic treatment plan. This may help in preventing any long-term complications associated with diabetes. Regular checking for blood sugars help in: 1) judging how well the patient is reaching the overall treatment plans and 2) understanding how diet and exercise affect the blood sugar levels. From Figure~\ref{fig_ps_1}, we see that patient ids corresponding to $\{260, 9, 24, 402, 435, 281, 139, 243\}$ are rarely checking their blood glucose levels. It can also be noticed that more than 90\% of the respondents are usually or always checking their blood glucose levels. 

\subsection{Insulin}

Insulin in the body helps in moving blood glucose from the blood stream into the cells for energy. However patients with diabetes cannot produce enough insulin. Hence they are required to take insulin injections regularly. Taking right dose of insulin at right time is necessary for the patients to lead an active and independent lives.

\begin{figure}[h]
\centering
\includegraphics[width=3.5in]{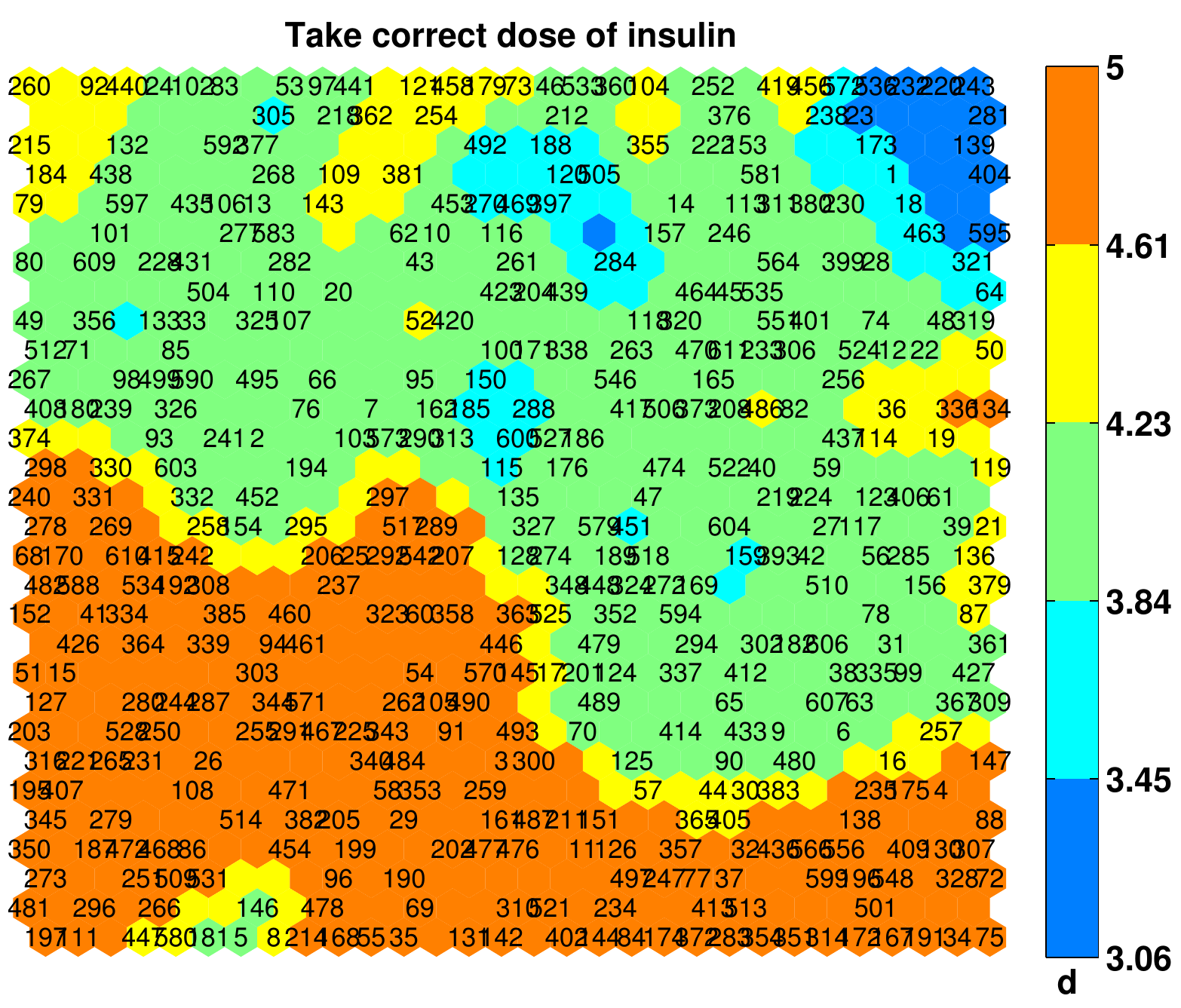}
\caption{Grouping for similar patients according to the ratings given for the self-care factor: `Take correct dose of insulin'.}
\label{fig_ps_4}
\end{figure}

\begin{figure}[h]
\centering
\includegraphics[width=3.5in]{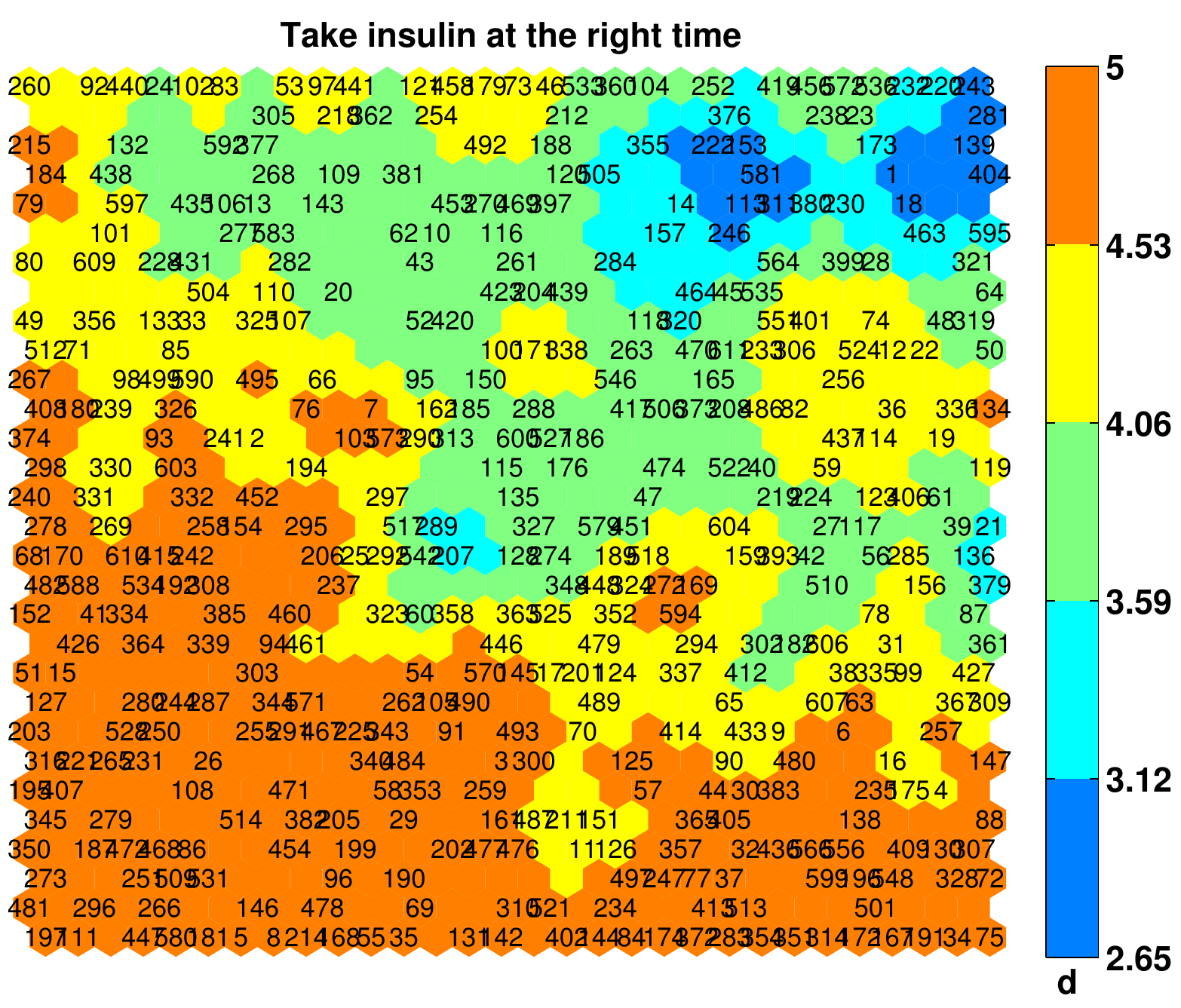}
\caption{Grouping for similar patients according to the ratings given for the self-care factor: `Take insulin at the right time'.}
\label{fig_ps_5}
\end{figure}

Patient ids corresponding to $\{243, 281, 139, 404\}$ sometimes take correct dose of insulin as shown in Figure~\ref{fig_ps_4}. It is interesting to notice that the same group of patients rarely take insulin at the right time as shown in Figure~\ref{fig_ps_5}. We also notice that patients taking correct dose of insulin also take insulin injections at the right time.  

\subsection{Food}

Eating of balanced food is an integral part of diabetes care and management. Diet control, balanced food selection, portion size and timely consumption of the food influences the diabetes treatment plan. Managing food as a self-care factor reduces the symptoms of diabetes by keeping the blood sugar levels and body weight under control.

\begin{figure}[h]
\centering
\includegraphics[width=3.5in]{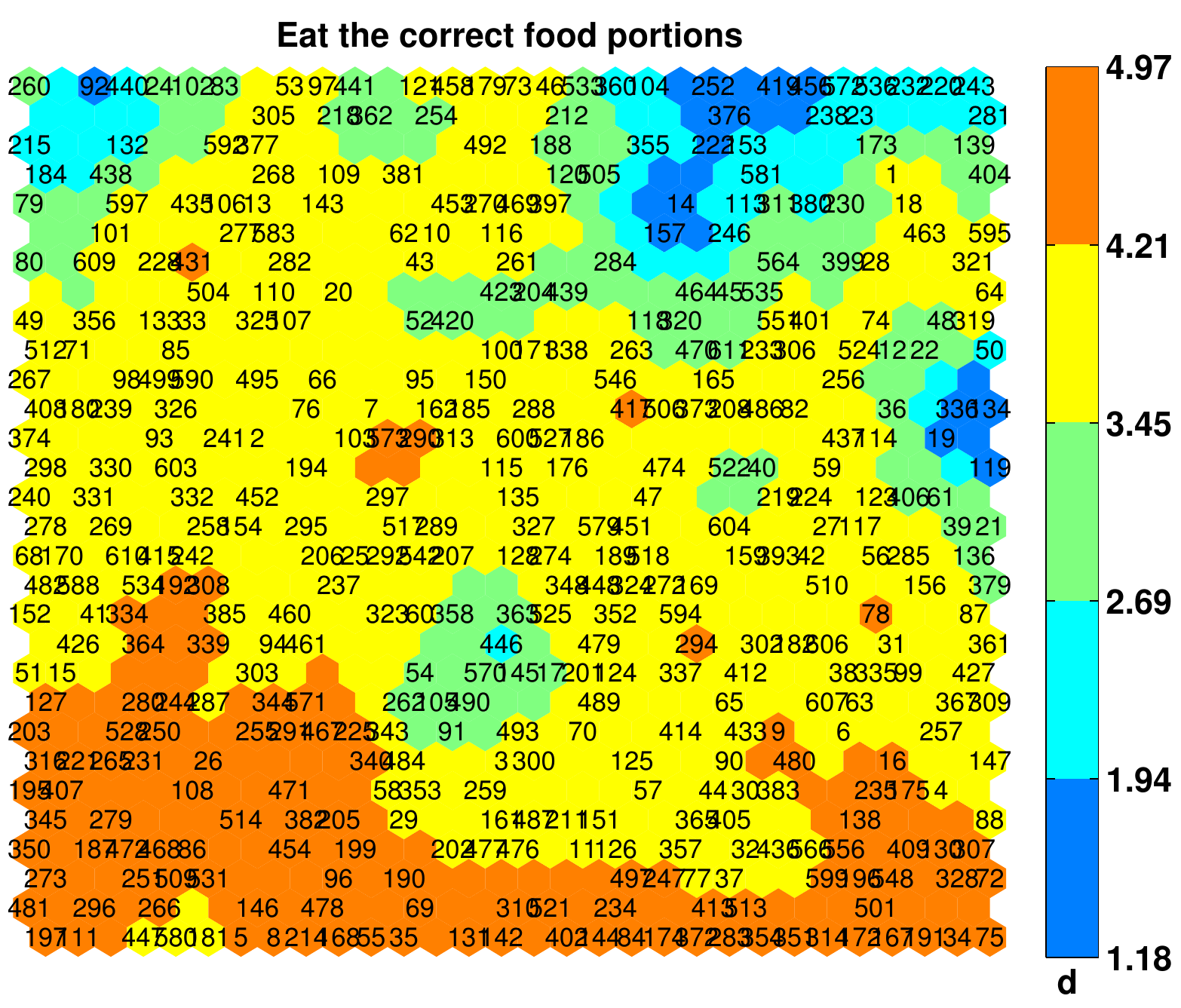}
\caption{Grouping for similar patients according to the ratings given for the self-care factor: `Eat the correct food portions'.}
\label{fig_ps_6}
\end{figure}

\begin{figure}[h]
\centering
\includegraphics[width=3.5in]{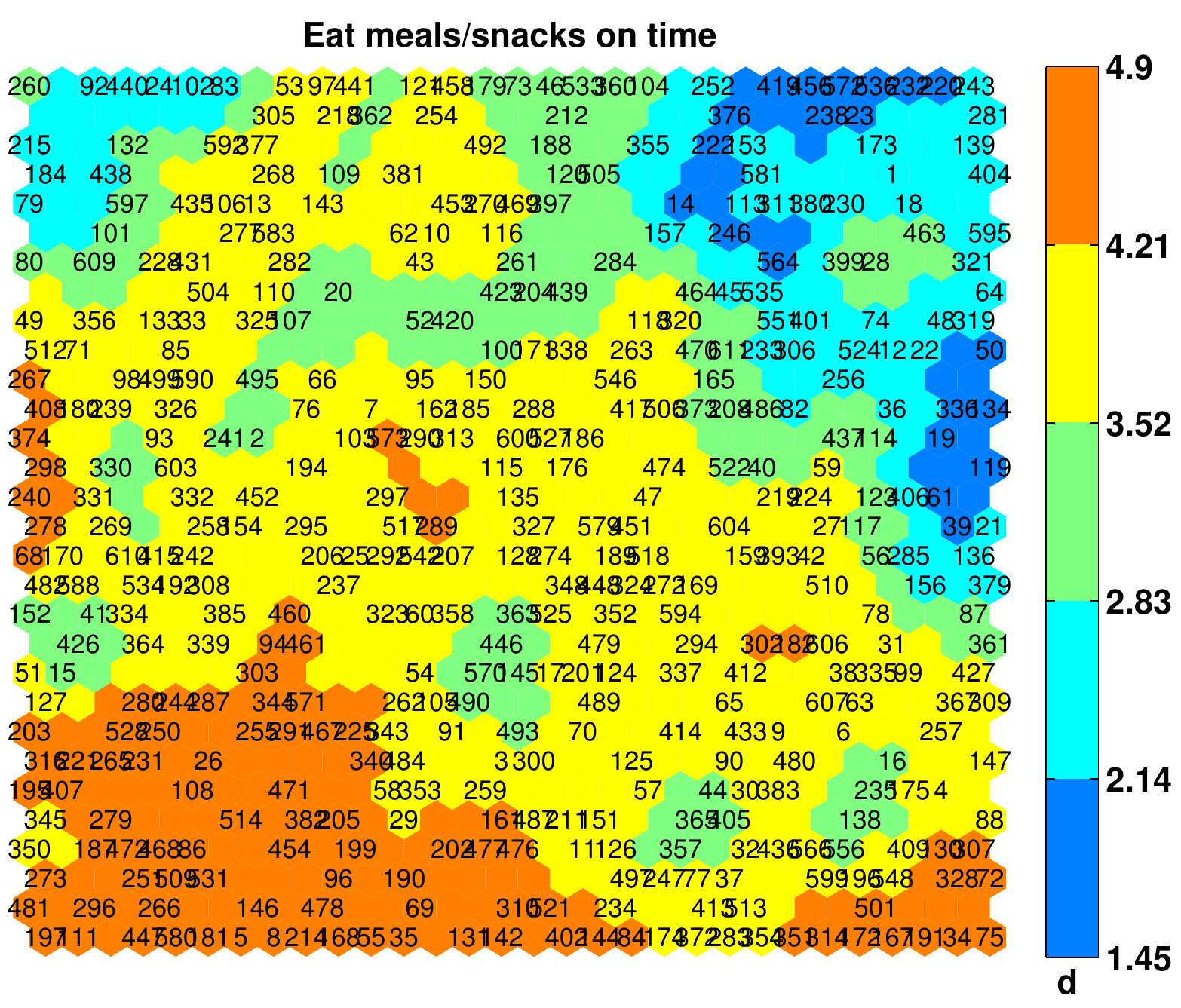}
\caption{Grouping for similar patients according to the ratings given for the self-care factor: `Eat meals on time'.}
\label{fig_ps_7}
\end{figure}

Figure~\ref{fig_ps_6} and Figure~\ref{fig_ps_7} have similar outlook revealing the correlation between them. From these figures, we see that patients who are not taking meals on time also do not take correct food portions. Similarly, patients eating on time take correct food portions. 

\subsection{Physical activity}

\begin{figure}[h]
\centering
\includegraphics[width=3.5in]{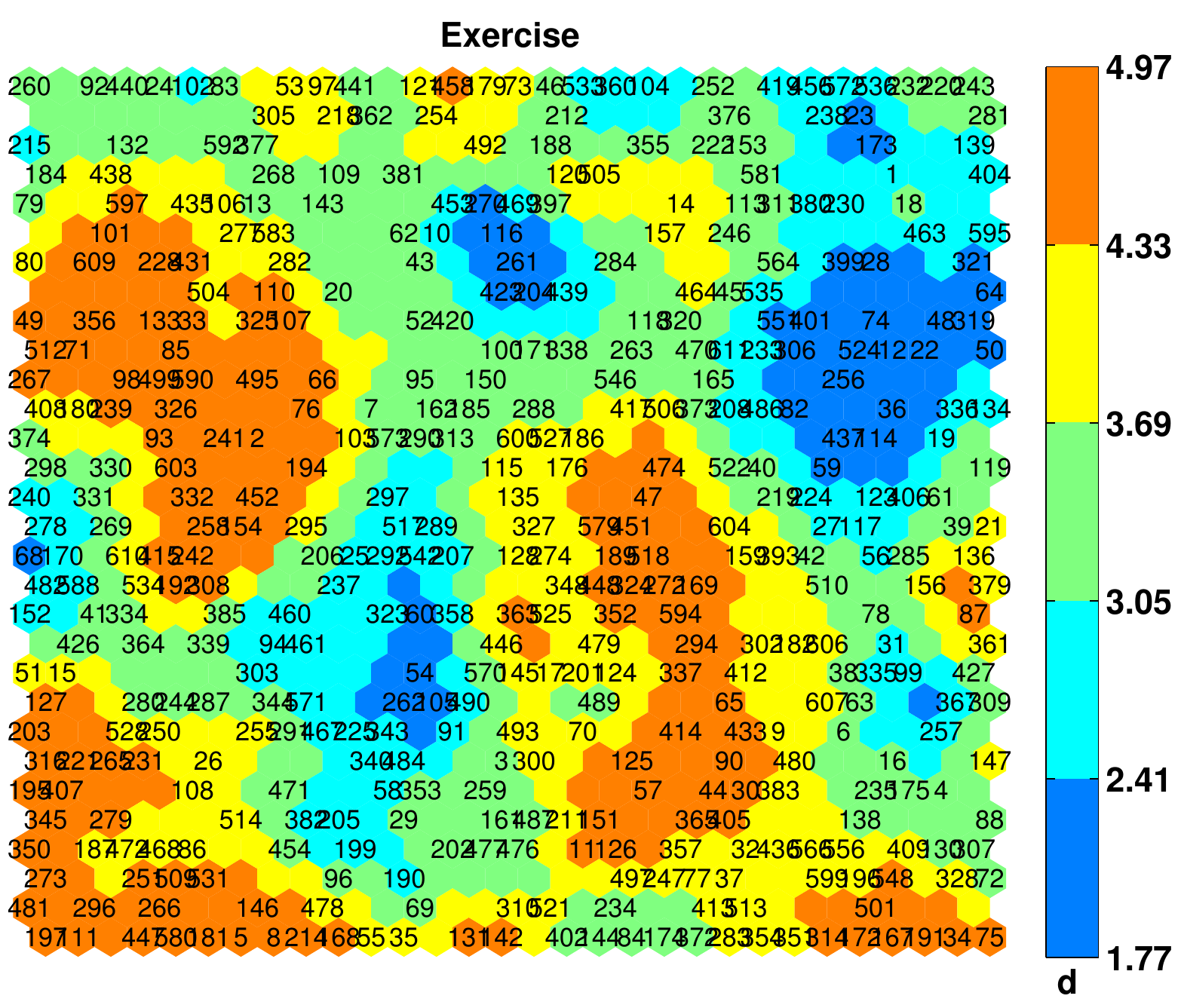}
\caption{Grouping for similar patients according to the ratings given for the self-care factor: `Exercise'.}
\label{fig_ps_14}
\end{figure}

Physical activity helps in keeping the weight under control. Keeping a healthy weight prevents diabetic complications. Exercise manages the blood sugars to stay in the target range.
It also helps the insulin to move glucose from the blood stream into the cells for energy. Hence, exercise can be considered a cornerstone of diabetes management, along with diet and medication.
Figure~\ref{fig_ps_14} shows the grouping of the patients according to their ratings for the factor 'Exercise'. It is seen that patients taking correct food portions perform regular physical activities.

\section{Conclusions and Discussions}
\label{condis}
This study shows the significance of SOM method
for inferring  useful information from the surveys. The results obtained using SOM analysis in this study provide a good understanding of the patient self-care behaviours. This study was also successful in identifying similar patient profiles that are in need of adjusting their lifestyles. The principle findings in this study show: 1) patients who take correct dose of insulin inject insulin at the right time, 2) patients who take correct food portions undertake regular physical activities and 3) patients who eat on time take correct food portions.

By attaining a better understanding of the patterns present in the behaviour of patients through machine learning, we are presented with the opportunity to classify patients who are in need of 
lifestyle adjustments in order to improve self care. Being able to target those who are most in need 
of assistance in developing life style changes will allow for improvements in the efficiency at which 
hospital resources are used. It can be difficult for the less affluent to attain self care tools in 
countries without national heath care~\cite{soltesz2009diabetes}. Identifying those who are most likely to be in need and making this apparent to clinicians can provide a more efficient allocation of hospital resources. This 
can result in more affordable self care tools for those who are likely to poorly manage their 
diabetes or provide free self care tools to those targeted as in need.


\section*{Acknowledgement}

The funding for this work has been provided by Department of Computing and Centre for Vision, Speech and Signal
Processing (CVSSP) - University of Surrey.



%
\bibliographystyle{IEEEtran}
\bibliography{santosh_references}

\end{document}